\documentclass[%
reprint,
superscriptaddress,
 amsmath,amssymb,
 aps,
prb,
]{revtex4-2}
\usepackage{graphicx}
\usepackage{dcolumn}
\usepackage{bm}

\usepackage{xcolor}
\usepackage{multirow}

\usepackage[normalem]{ulem}
\newcommand{\editor}[2]{%
  \expandafter\newcommand\csname #1note\endcsname[1]{%
    \textcolour{#2}{(\textbf{#1:} \it ##1)}}%
  \expandafter\newcommand\csname #1\endcsname[1]{%
    \textcolour{#2}{##1}}%
  \expandafter\newcommand\csname #1cancel\endcsname[1]{%
    \textcolour{#2}{\sout{##1}}}%
  \expandafter\newcommand\csname #1change\endcsname[2]{%
    \textcolour{#2}{\sout{##1} ##2}}%
  \newenvironment{#1text}{\colour{#2}}{\colour{black}}
}

\editor{FP}{orange}
\editor{CS}{blue}

\begin{document}
\preprint{APS/123-QED}
\title{Valley-Resolved Strain Engineering of Excitons in Monolayer Transition Metal Dichalcogenides}

\author{Cem Sevik}
\affiliation{COMMIT, Department of Physics and NANOlight Center of Excellence, University of Antwerp, Groenenborgerlaan 171, B-2020 Antwerp, Belgium}
\author{Purushothaman Manivannan} 
\affiliation{COMMIT, Department of Physics and NANOlight Center of Excellence, University of Antwerp, Groenenborgerlaan 171, B-2020 Antwerp, Belgium}
\author{Fulvio Paleari}
\affiliation{Centro S3, CNR-Istituto Nanoscienze, I-41125 Modena, Italy}
\author{Milorad V. Milo\v{s}evi\'c}
\email{milorad.milosevic@uantwerpen.be; mmilosev@nd.edu}
\affiliation{COMMIT, Department of Physics and NANOlight Center of Excellence, University of Antwerp, Groenenborgerlaan 171, B-2020 Antwerp, Belgium}
\affiliation{Stavropoulos Center for Complex Quantum Matter, University of Notre Dame, Notre Dame, IN, USA}

\date{\today}

\begin{abstract}
We present a comprehensive many-body investigation of strain-dependent excitonic properties in monolayer transition metal dichalcogenides, including MoS$_2$, MoSe$_2$, WS$_2$, and WSe$_2$. Using fully relativistic and tightly converged GW and Bethe-Salpeter equation calculations, we systematically analyze the evolution of both direct ($\mathbf{Q}$ = 0) and indirect ($\mathbf{Q}$ $\neq$ 0) excitonic states under biaxial strain. Particular emphasis is placed on identifying the momentum-space character and valley composition of excitons through reciprocal-space excitonic weight analysis, enabling a detailed classification of intervalley and intravalley excitonic states beyond conventional optical peak assignments. We show that strain strongly modifies the energetic ordering and hybridization of excitons associated with the K, $\Gamma$, and $\Lambda$ valleys, leading to distinct and, in some cases, opposite strain responses for different indirect excitons. Quantitative strain gauge factors are extracted for both direct and indirect excitonic transitions and compared with available photoluminescence, angle-resolved photoemission spectroscopy, magneto-optical, and electron energy loss spectroscopy measurements. Our calculations reproduce the experimentally observed qualitative trends of strain-induced excitonic shifts and provide a microscopic interpretation of their origin in terms of valley-dependent quasiparticle band-structure evolution. Our results particularly highlight the importance of many-body-corrected electronic structures for reliable exciton engineering in two-dimensional semiconductors.
\end{abstract}

\maketitle

\section{Introduction}
The emergence of atomically thin two-dimensional (2D) materials has enabled the investigation of electronic and optical phenomena in the limit of reduced dimensionality and screening \cite{2d_app-1,2d_app-2}. Among these systems, monolayer transition metal dichalcogenides (TMDs), such as MoS$_2$, MoSe$_2$, WS$_2$, and WSe$_2$, have attracted particular attention due to their direct band gaps at the K valleys and the strong Coulomb interactions originating from weak dielectric screening \cite{mos2-gapsAl,band-gaps,band-gaps2,mos2-gaps2,MoS2_exp1}. In addition, strong spin-orbit coupling together with broken inversion symmetry leads to coupled spin and valley degrees of freedom, making monolayer TMDs an important platform for studying valley-dependent optical phenomena \cite{FP-Ref1}. The reduced dimensionality and enhanced electron-hole interactions further result in strongly bound excitons with binding energies on the order of several hundred meV, causing excitonic effects to dominate the optical response even at room temperature \cite{mos2-pl1,mos2-pl3,mos2-pl4,mos2-pl5,mos2-pl6,mos2-pl7,mos2-pl8,mos2-pl9,mos2-pl10,MA-1}.

The strong excitonic response of monolayer TMDs has motivated extensive experimental and theoretical investigations into their many-body optical properties. In addition to bright neutral excitons, experiments have identified a variety of excitonic complexes including trions, biexcitons, dark excitons, and intervalley excitons \cite{mak2013tightly,you2015observation,zhang2015experimental,rivera2015observation}. Valley-selective optical selection rules further enable the generation and manipulation of valley polarization and coherence through circularly polarized light \cite{FP-Ref1,mak2012control}. More recently, the excitonic landscape of TMD-based heterostructures has been further enriched by the emergence of moiré and interlayer excitonic phenomena in twisted van der Waals systems \cite{tran2019evidence,wang2020correlated,liu2015strong}. These developments have established monolayer TMDs as a versatile platform for investigating many-body quantum phenomena and exciton-based optoelectronic functionalities.

Considerable effort has therefore been devoted to engineering excitonic properties in monolayer TMDs using external perturbations such as electrostatic gating, dielectric environment control, and mechanical deformation \cite{chernikov2015electrical,mos2-pl4,mos2-pl16,he2013experimental,stier2016exciton}. Among these approaches, strain engineering has emerged as a particularly effective method for tuning both the electronic band structure and excitonic landscape of atomically thin semiconductors. Biaxial and uniaxial strain can modify the relative energy positions of the $\Gamma$, K, and $\Lambda=\frac{1}{2}|\textbf{K}|$ valleys, alter effective masses, and influence dielectric screening, leading to substantial changes in excitonic transition energies and direct–indirect excitonic behavior \cite{mos2-pl12,mos2-pl14,mos2-pl15,mos2-pl19,javed2024band,Eb_MoS2_WS2_Shi-2013}. Experimentally, strain-dependent photoluminescence and absorption measurements have demonstrated that moderate tensile strain of approximately 1–2\% can shift bright excitonic resonances by up to 100 meV and induce the optical activation of otherwise dark excitonic states \cite{mos2-pl20,Eb_MoS2_Yu-2019}. Complementary theoretical studies based on density functional theory, effective low-energy models, and many-body approaches have further shown that strain can significantly modify quasiparticle dispersions, valley ordering, and excitonic properties through changes in orbital hybridization and screening effects \cite{Eb_MoS2_WS2_Shi-2013,mos2-pl15,Deilmann2019,emalik-1,Strain-1, scalise2012strain,yun2012thickness}.

Despite these advances, a unified many-body description of the strain-dependent excitonic landscape of monolayer TMDs remains incomplete. Previous GW+BSE studies have predominantly focused on optically bright, zero-momentum excitons, whereas the evolution of finite-momentum excitons under strain has received considerably less attention. This is particularly important in multivalley TMDs, where relatively small strain-induced shifts of the electronic states at K, $\Gamma$, and $\Lambda$ can modify not only exciton energies but also the ordering, valley character, and hybridization of low-lying excitonic states. Consequently, understanding strain-dependent exciton physics requires going beyond conventional optical peak assignments and explicitly resolving the momentum-space composition of the excitonic wave functions. An additional open question is to what extent these excitonic responses can be traced back to the strain evolution of the underlying quasiparticle valleys, whose relative energies can be substantially modified by many-body corrections.

Here, we address these questions through a systematic GW+BSE investigation of strained monolayer MoS$_2$, MoSe$_2$, WS$_2$, and WSe$_2$, including both zero- and finite-center-of-mass-momentum excitons. By analyzing the reciprocal-space weights of the BSE eigenstates, we resolve their valley composition and follow the evolution of individual excitonic states under biaxial strain. We show that strain reorganizes the low-energy excitonic manifold in a strongly valley-selective manner: excitons associated with different valley configurations can exhibit opposite strain responses, undergo changes in energetic ordering, and develop substantial hybridization. We further demonstrate that these trends closely follow the strain-dependent evolution of the corresponding quasiparticle valley gaps and that many-body corrections substantially modify the resulting strain gauge factors compared with the underlying DFT description. Comparison with available ARPES, photoluminescence, magneto-optical, and EELS measurements provides experimental context for these trends. Our results establish a microscopic connection between strain-induced quasiparticle valley reconstruction and the resulting finite-momentum excitonic landscape, providing a many-body framework for valley-selective exciton engineering in monolayer TMDs.

\section{Results}
\subsection{Electronic properties: Strain-free}
The relative energies of the electronic states at K, $\Gamma$, and $\Lambda$ provide the single-particle framework underlying the low-energy excitonic landscape of monolayer TMDs. In particular, we consider the valence-band offset \(\Delta E^{v}_{\mathrm{K}\Gamma}=E^{v}_{\mathrm{K}}-E^{v}_{\Gamma}\) and the conduction-band offset \(\Delta E^{c}_{\mathrm{K}\Lambda}=E^{c}_{\mathrm{K}}-E^{c}_{\Lambda}\). Together with the direct K–K gap, these quantities determine the energetic proximity of the electronic transitions from which the relevant zero- and finite-momentum excitons originate. Their accurate description is therefore essential for interpreting the valley-resolved excitonic states discussed below.

\begin{figure}[t]
\includegraphics[width=\linewidth]{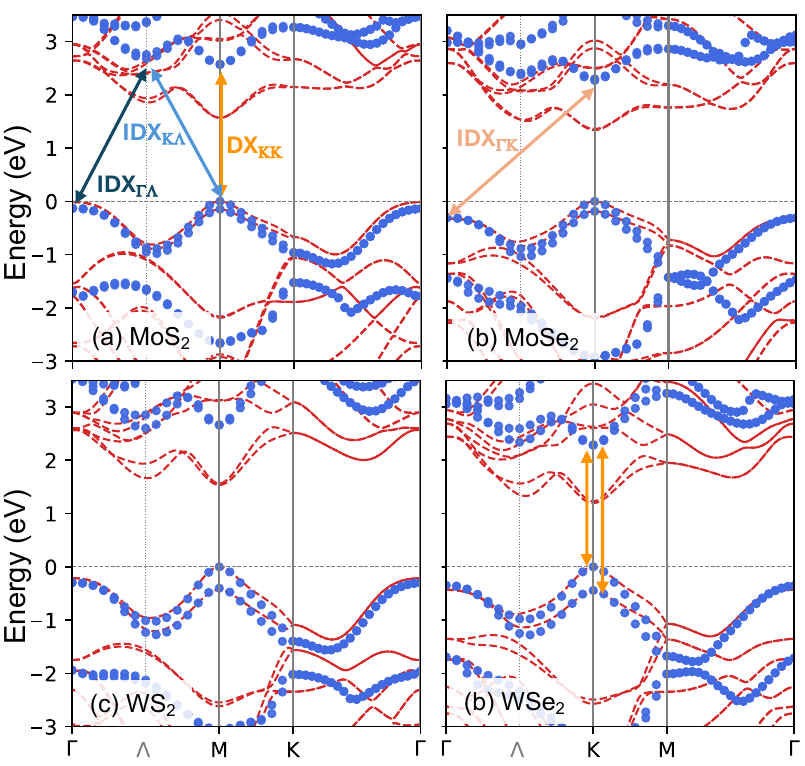}
\caption{\label{fig-band} The calculated PBE (dashed red lines) and G$_0$W$_0$ (blue symbols) band structure of (a) MoS$_2$, (b) MoSe$_2$, (c) WS$_2$, (d) WSe$_2$.}
\end{figure}

Figure \ref{fig-band} compares the PBE and \(G_0W_0\) band structures of the four monolayers, while the corresponding valley offsets are summarized in Table I. Importantly, the many-body corrections are strongly valley dependent rather than acting as a simple rigid shift of the conduction and valence bands. For example, \(\Delta E^{c}_{\mathrm{K}\Lambda}\) changes by more than 150 meV between PBE and \(G_0W_0\) for the S-based compounds and by approximately 85 meV for the Se-based compounds. The valence-band offset \(\Delta E^{v}_{\mathrm{K}\Gamma}\) is likewise substantially modified. These changes alter the relative energies of the K–K, K–$\Lambda$, $\Gamma$–K, and $\Gamma$–$\Lambda$ single-particle transitions and therefore reshape the energetic landscape from which the corresponding excitonic states emerge. Collectively, these observations highlight the critical importance of many-body corrections for accurately describing the valley-resolved electronic structure that underlies both direct and finite-momentum excitons, consistent with previous studies \cite{ARPES16, Wang2018, Kunstmann2018,Eb_MoS2_MoSe2_WS2_WSe2_Rama-2012}.

Strikingly, G$_0$W$_0$ correction induce appreciable shifts not only in the conduction band states but also in the deeper valence states, which are primarily composed of chalcogen $p$-orbitals~\cite{Eb_MoS2_WS2_Shi-2013,Debbichi2014}. While these modifications do not directly influence the low-energy excitonic states of primary interest, they underscore the broader impact of many-body interactions on the electronic and optical properties of TMDs, including those with metallic character.

In this context, computational accuracy is particularly important. Quantities such as quasiparticle gaps and relative valley offsets are sensitive to the choice of pseudopotentials, plane-wave cutoffs, $k$-point sampling, and other numerical parameters entering both (G$_0$W$_0$) and BSE calculations. Indeed, previously reported DFT (GW) band gaps span 1.58-1.89 eV (2.41-2.82 eV), 1.35-1.63 eV (2.31-2.41 eV), 1.56-2.10 eV (2.30-2.80 eV), and 1.19-1.74 eV (2.42-2.51 eV) for MoS$_2$, MoSe$_2$, WS$_2$, and WSe$_2$, respectively~\cite{Rafael, javed2024band, rodrigues2024critical}. Such variations are particularly relevant here because the energy separations among competing low-lying excitonic states are typically only of the order of a few hundred meV or less. Even considerably smaller errors in the relative valley energies can therefore modify the predicted ordering and character of finite-momentum excitons. To minimize these uncertainties, we employ fully relativistic, norm-conserving pseudopotentials and tightly converged numerical parameters, with all relevant quantities subjected to systematic convergence tests.

\begin{table}[ht!]
\caption{\label{table1}The calculated electronic properties of the considered materials. The presented values correspond to the direct band gap, indirect band-gap, $\mathrm{K}-\Lambda$ energy difference at lowest conduction levels, $\mathrm{K}-\Gamma$ energy difference at highest valence levels, and spin-orbit valence splitting at K/K$^\prime$, respectively, as defined in Figure \ref{fig-band}. Here, the previously undefined acronyms refer to the following: LDA stands for Local Density Approximation, GGA for Generalized Gradient Approximation, and FP for Full Potential. }
\begin{ruledtabular}
\begin{tabular}{llcccccr}
TMD&MTD&$\Delta E_{\mathrm{KK  }}$&$\Delta E_{\mathrm{K\Lambda}}$&$\Delta E_{\mathrm{K\Gamma}}^{v}$&$\Delta E_{\mathrm{K\Lambda}}^{c}$& $\Delta_{\mathrm{SOC}}$&\\
&&(eV)&(eV)&(meV)&(meV)& (meV)&Ref\\\hline
\multirow{8}{*}{MoS$_2$} &PBE&\textbf{1.566}&1.859&11&-293&149&TW\\
&G$_0$W$_0$&\textbf{2.566}&2.678&131&-113&149&TW\\
&PBE&1.690&&20&-250&&\cite{band-gaps2}\\
&G$_0$W$_0$&2.750&&160&-230&&\cite{band-gaps2}\\
&G$_0$W$_0$&2.780&&&&146&\cite{qiu2013optical}\\
&G$_1$W$_0$&2.840&&&&&\cite{qiu2013optical}\\
&QSGW&2.760&&&&&\cite{cheiwchanchamnangij}\\
&LDA&&&145&-107&&\cite{emalik-1}\\
&FP&1.790&&190&-50&&\cite{mos2-gapsAl}\\
&GGA&1.710&&210&&150&\cite{mos2-gaps2}\\
&FP&&&50&&148&\cite{band-gaps}\\\hline
\multirow{6}{*}{MoSe$_2$} &PBE&\textbf{1.339}&1.530&323&-191&185&TW\\
&G$_0$W$_0$&\textbf{2.269}&2.374&306&-105&188&TW\\
&PBE&1.430&&230&-230&&\cite{band-gaps2}\\
&G$_0$W$_0$&2.330&&340&-330&&\cite{band-gaps2}\\
&LDA&&&317&-171&&\cite{emalik-1}\\
&FP&&&360&&183&\cite{band-gaps}\\\hline
\multirow{6}{*}{WS$_2$} &PBE&\textbf{1.537}&1.665&212&-128&430&TW\\
&G$_0$W$_0$&2.661&\textbf{2.602}&294&59&401&TW\\
&PBE&1.780&&$<$2&-250&&\cite{band-gaps2}\\
&G$_0$W$_0$&2.880&&60&-250&&\cite{band-gaps2}\\
&LDA&&&237&-55&&\cite{emalik-1}\\
&FP&&&220&&426&\cite{band-gaps}\\\hline
\multirow{6}{*}{WSe$_2$} &PBE&\textbf{1.198}&1.352&430&-154&471&TW\\
&G$_0$W$_0$&\textbf{2.291}&2.341&347&-50&442&TW\\
&PBE&1.500&&260&-210&&\cite{band-gaps2}\\
&G$_0$W$_0$&2.380&&340&-360&&\cite{band-gaps2}\\
&LDA&&&484&-1&&\cite{emalik-1}\\
&FP&&&530&&456&\cite{band-gaps}\\
\end{tabular}
\end{ruledtabular}
\end{table}

To further assess the accuracy of our computational framework, we compare our results with available experimental data. Angle-resolved photoemission spectroscopy (ARPES), in particular, provides high-resolution measurements of the highest occupied valence states and spin-orbit splittings. The experimentally reported values of $\Delta E^v_{\mathrm{K\Gamma}}$, summarized in Table~\ref{table2}, show considerable variability—even within similar sample types—likely arising from structural and environmental factors such as substrate-induced dielectric screening, residual strain, and variations in growth conditions\cite{mos2-pl26,ARPES5}. Despite these discrepancies, the consistent identification of the K point as the valence band maximum across all measurements reinforces the reliability of our theoretical predictions.

\begin{table}[ht!]
\caption{\label{table2} The experimental results for the band structure of the considered materials. Here, Ex, CVD, and MBE denote the growth techniques used to prepare the samples—Exfoliation, Chemical Vapor Deposition, and Molecular Beam Epitaxy, respectively. The quantities $\Delta E$, SOC, $X_{B}$, $\Delta E^{v}_{\mathrm{K\Gamma}}$, and $T$ correspond to the estimated band gap, the spin–orbit–induced valence-band splitting at K/K$^\prime$, the binding energy of the direct exciton, the energy separation between the K and $\Gamma$ points at the valence-band maximum, and the photoluminescence measurement temperature, respectively. SUB denotes the substrate.}
\begin{ruledtabular}
\begin{tabular}{lllcccccr}
TMD&EXP&SUB&$\Delta E$&SOC&X$_{B}$&$\Delta^{v}_{\mathrm{K}\Gamma}$&T&REF\\
&&&(eV)&(meV)&(meV)&(meV)&(K)&\\\hline
\multirow{4}{*}{MoS$_2$}&Ex&$h$-BN&2.07&170& &140&100&[\onlinecite{ARPES2}]\\
&CVD&Sp&2.11& &240&20&300&[\onlinecite{ARPES5}]\\
&CVD&Au&1.90& &90&60&300&[\onlinecite{ARPES5}]\\
&CVD&Gr&&&&120&11&[\onlinecite{ARPES20}]\\\hline
\multirow{4}{*}{MoSe$_2$}&Ex&BN&1.64&220&&480&100&[\onlinecite{ARPES2}]\\
&MBE&Gr&2.15&240&&390&110&[\onlinecite{ARPES6}]\\
&Ex&Gr&2.18&&550&&&[\onlinecite{ARPES16}]\\
&Ex&Gr&&240&&440&110&[\onlinecite{ARPES4}]\\\hline
\multirow{7}{*}{WS$_2$}&CVD&Gr&2.47&425&&240&300&[\onlinecite{ARPES14}]\\
&Ex&BN&2.03&450&&390&100&[\onlinecite{ARPES2}]\\
&CVD&Gr&&440&&200&8&[\onlinecite{ARPES11}]\\
&CVD&Gr&&462&&182&300&[\onlinecite{ARPES12}]\\
&CVD&TiO$_2$&&420&&260&85&[\onlinecite{ARPES15}]\\
&CVD&Gr&&420&&200&300&[\onlinecite{ARPES17}]\\
&CVD&Si&2.04&&&150&&[\onlinecite{ARPES16}]\\\hline
\multirow{9}{*}{WSe$_2$}&MBE&Gr&1.95&475&210&560&60&[\onlinecite{ARPES1}]\\
&MBE&Gr&2.12&440&500&640&77&[\onlinecite{ARPES6}]\\
&Ex&BN&1.79&485&&620&100&[\onlinecite{ARPES2}]\\
&CVD&Sp&1.89&&240&260&300&[\onlinecite{ARPES5}]\\
&CVD&Au&1.75&&140&200&300&[\onlinecite{ARPES5}]\\
&Ex&Au&&460&&680&30&[\onlinecite{ARPES3}]\\
&Ex&Gr&&490&&500&110&[\onlinecite{ARPES4}]\\
&Ex&Si&&513&&892&300&[\onlinecite{ARPES7}]\\
&CVD&Gr&&500&&400&300&[\onlinecite{ARPES8}]\\
\end{tabular}
\end{ruledtabular}
\end{table}

Moreover, our G$_0$W$_0$ calculations reproduce $\Delta E^v_{\mathrm{K}\Gamma}$ values that are in close agreement with experimental results obtained from monolayers supported on graphene or encapsulated in hexagonal boron nitride ($h$-BN). The calculated spin-orbit splittings also show excellent agreement with experimental averages, with a slight systematic underestimation that remains within typical experimental uncertainty. In contrast, the G$_0$W$_0$ band gaps obtained in this study tend to exceed experimental values, which we attribute to extrinsic effects such as sample charging and band bending during optical or photoemission measurements, together with band gap reduction due to electron-phonon coupling. Nonetheless, the resulting direct exciton energies derived from our calculations are in very good agreement with experimental optical spectra, as discussed in the subsequent sections. Taken together, there is a robust correspondence between our highly converged G$_0$W$_0$ results and experimental benchmarks for $\Delta E^v_{\mathrm{K}\Gamma}$ and spin-orbit splitting of monolayer TMDs.

\begin{figure}[ht!]
\includegraphics[width=\linewidth]{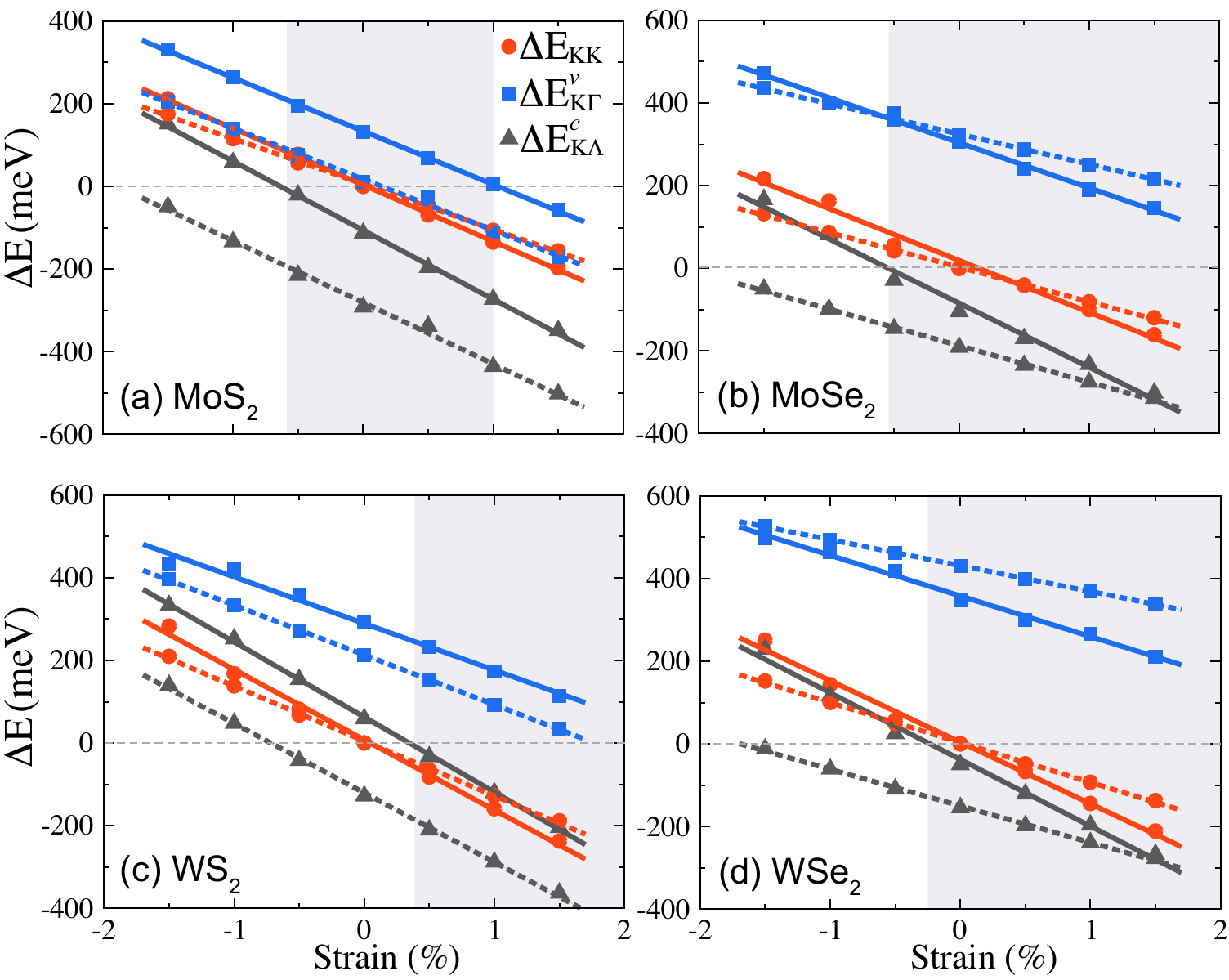}
\caption{\label{fig-strain}Calculated PBE (dashed lines) and G$_{0}$W$_{0}$ (solid lines) band-gap transitions under biaxial strain for (a) MoS$_2$, (b) MoSe$_2$, (c) WS$_2$, and (d) WSe$_2$. The quantities $\Delta_{\mathrm{KK}}$, $\Delta^{v}_{\mathrm{K\Gamma}}$, and $\Delta^{c}_{\mathrm{K\Lambda}}$ denote, respectively, the strain-induced change in the direct band gap at K (red circles), the energy difference between the highest valence states at K and $\Gamma$ (blue squares), and the energy separation between the lowest conduction states at K and $\Lambda$ (black triangles). The gray shaded area represents the strain windows where the band gap is direct at K.} 
\end{figure}

\subsection{Electronic properties: Strained}
Strain engineering is well established as a powerful strategy for tailoring the electronic band structure of monolayer TMDs. Both theoretical and experimental studies have shown that uniaxial and biaxial strain modify the relative energies of the K, $\Lambda$, and $\Gamma$ valleys, often driving transitions between direct and indirect band-gap configurations. First-principles and many-body calculations further show that these valley shifts strongly affect the energy separation between bright and momentum-forbidden excitonic states \cite{Strain-1, emalik-1}. These trends are consistent with strain-dependent photoluminescence and ARPES measurements, which reveal redshifts of the direct gap together with material-specific changes in dark-valley states \cite{emalik-2, Strain-2}. In particular, recent strain-resolved photoluminescence measurements have introduced exciton strain gauge factors to quantify valley-specific responses and have revealed opposite strain-induced trends for K–K and K–$\Lambda$ transitions in WSe$_2$ \cite{mos2-pl15}. These observations underscore the strongly valley-dependent character of the strain response in monolayer TMDs.


The substantial strain-induced changes in $\Delta E^{v}_{\mathrm{K}\Gamma}$ and $\Delta E^{c}_{\mathrm{K}\Lambda}$ directly affect the relative ordering of the electronic transitions underlying zero-momentum (direct) and finite-momentum (indirect) excitonic states, even for modest strain values. Within the G$_0$W$_0$ description, the valence-band maximum (VBM) remains at the K point throughout the considered strain range for all materials except MoS$_2$, as indicated by the evolution of $\Delta E^{v}_{\mathrm{K}\Gamma}$ in Fig.~\ref{fig-strain}. For MoS$_2$, compressive strain beyond approximately 1.0\% shifts the VBM from K to $\Gamma$ at the G$_0$W$_0$ level, whereas the corresponding transition already occurs at about 0.1\% within PBE. Similar K-to-$\Gamma$ valence-band crossovers are predicted at approximately 2.8\% (4.4\%), 2.6\% (1.8\%), and 3.6\% (7.0\%) strain for MoSe$_2$, WS$_2$, and WSe$_2$, respectively, where the values in parentheses correspond to PBE.

An analogous analysis can be performed for the conduction-band offset $\Delta E^{c}_{\mathrm{K}\Lambda}$. A direct band gap at K requires the appropriate ordering of both the valence and conduction valleys, namely a positive $\Delta E^{v}_{\mathrm{K}\Gamma}$ together with a negative $\Delta E^{c}_{\mathrm{K}\Lambda}$. The corresponding strain windows are indicated by the shaded regions in Fig.~\ref{fig-strain}. Notably, the range of strain over which the band gap remains direct is narrower for the S-based compounds than for the Se-based monolayers, underscoring their stronger sensitivity to strain-induced valley reordering. Considering both the G$_0$W$_0$ results and the uncertainty associated with DFT-predicted equilibrium lattice constants, the precise band-gap character near zero strain remains somewhat delicate, although a direct gap is more robustly favored in the Se-based monolayers.

\begin{table}[ht!]
\caption{\label{table3} Exciton peak energy A (eV) and binding energy E$_b$ (meV) for various 2D materials from this work and literature. Here, TW, WE, and GW refer to the calculations performed in this work, those obtained from the Wannier equation, and those computed using the GW-BSE method, respectively.}
\begin{ruledtabular}
\begin{tabular}{cccccccccc}
\multicolumn{2}{c}{MoS$_2$}&\multicolumn{2}{c}{MoSe$_2$}&\multicolumn{2}{c}{WS$_2$}&\multicolumn{2}{c}{WSe$_2$}& Method & Ref.  \\
E$_{\mathrm{A}}$& E$_b$ &E$_{\mathrm{A}}$& E$_b$ &E$_{\mathrm{A}}$& E$_b$ &E$_{\mathrm{A}}$& E$_b$& &\\
\hline
{1.94}&{ 625}&{1.70}&{570}&2.08&{ 580}&1.75& 530&TW & \\
{1.97}&{ 850}&{1.66}&{750}&2.17&{ 710}&1.75&670& GW &\cite{Eb_MoS2_MoSe2_WS2_WSe2_Rama-2012} \\
{1.78}&{1030}&{1.50}&{910}&1.84&{1040}&1.52&900& WE &\cite{Eb_MoS2_MoSe2_WS2_WSe2_Rama-2012} \\
{1.87}&{1020}&-&-&1.97&1050&-&-& GW &\cite{Eb_MoS2_WS2_Shi-2013} \\
{2.22}&{ 540}&&&2.51&{ 540}&&& GW &\cite{Eb_MoS2_WS2_Shi-2013} \\
{2.04}&{ 630}&-&-&-&-&-&-& GW &\cite{Eb_MoS2_qiu_Qiu-2016} \\
{1.80}&{ 620}&&&&&&& GW &\cite{Eb_MoS2_Yu-2019} \\
{1.88}&{ 960}&&&&&&& GW &\cite{qiu2013optical} \\
{1.98}&{-}&-&-&-&-&{1.65}&-& TD-DFT &\cite{Eb_MoS2_WSe2-Tomi-2020} \\
{-}&{ 673}&&{648}&&{610}&& 533& GW & \cite{Eb_MoS2_MoSe2_WS2_WSe2-Spinorial-2021} \\
{-}&{-}&{1.62}&{510}&-&-&-&-& GW &\cite{Eb_MoSe2_G_Wang-2015_2D} \\
{-}&{-}&{-}&{-}&-&-&1.69&394& GW &\cite{Eb_WSe2_Shih-2024} \\
\end{tabular}
\end{ruledtabular}
\end{table}

\subsection{Excitonic properties: Strain-free}\label{sec:BSEresults}
As discussed above, many-body corrections significantly affect the key quasiparticle energy levels that govern the optical response of monolayer TMDs. To examine how these corrections influence excitonic features, we solved the Bethe–Salpeter equation on top of G$_0$W$_0$ quasiparticle energies using the Yambo code. The lowest-energy bright exciton in all four materials is governed predominantly by electron–hole interactions at the K point, labeled as DX$_{\mathrm{KK}}$ in Fig. \ref{fig-band}(a).
This state, denoted as the A-exciton in the literature, exhibits excitation energies of 1.94, 1.70, 2.08, and 1.75 eV for monolayer MoS$_2$, MoSe$_2$, WS$_2$, and WSe$_2$, respectively. These values show excellent agreement with low-temperature photoluminescence measurements, which report A-exciton peaks in the ranges of 1.90–1.95 eV for MoS$_2$ \cite{mos2-pl1, mos2-pl3, mos2-pl8, mos2-pl9, mos2-pl10}, 1.64–1.66 eV for MoSe$_2$ \cite{mos2-pl3, mos2-pl4, mos2-pl5, mos2-pl6, mos2-pl7, mos2-pl24, mos2-pl25}, 2.08–2.12 eV for WS$_2$ \cite{mos2-pl12, mos2-pl15, mos2-pl16, mos2-pl17, mos2-pl19}, and 1.72–1.75 eV for WSe$_2$ \cite{mos2-pl3, mos2-pl13, mos2-pl14, mos2-pl15, mos2-pl18, mos2-pl20, mos2-pl21, mos2-pl22, mos2-pl23, mos2-pl24, mos2-pl26}. This close correspondence underscores the predictive accuracy of our G$_0$W$_0$+BSE approach for capturing the fundamental excitonic properties of these monolayers. These results are summarized in Table~\ref{table3}, along with previously published GW-BSE values. The corresponding A-exciton energies obtained in our calculations fall within the spread of earlier theoretical reports: 1.78–2.22 eV for MoS$_2$, 1.50–1.66 eV for MoSe$_2$, 1.84–2.51 eV for WS$_2$, and 1.52–1.75 eV for WSe$_2$. 

Furthermore, our calculations accurately reproduce the A–B exciton energy splitting arising from spin–orbit coupling in the valence band. We obtain splittings of 144, 195, and 358 meV for MoS$_2$, MoSe$_2$, and WS$_2$, respectively, which compare favourably with the experimentally reported values of approximately 150 meV \cite{mos2-pl1, mos2-pl10, mos2-pl11}, 200 meV, and 380 meV \cite{mos2-pl17, mos2-pl11}.

Exciton binding energies (E$_b$) reported in the literature span a broad range, largely due to variations in sample quality, measurement techniques, and environmental screening. For example, experimental E$_b$ values for A-exciton range from 217 to 640 meV for monolayer MoS$_2$  \cite{MA-1, Mos2-pl27, Mos2-pl29, STS-1, MoS2-pl28, Eb_exp-1}, 216 to 590 meV for MoSe$_2$ \cite{Mos2-pl27, MA-1, ARPES6, ARPES16, mos2-pl24, Eb_exp-1}, 174 to 700 meV for WS$_2$ \cite{MA-1, Mos2-pl27, mos2-pl11, STS-1, Eb_exp-1}, 167 to 720 meV for WSe$_2$ \cite{Mos2-pl27, MA-1, Mos2-pl29, mos2-pl26, Eb_exp-1}. Our G$_0$W$_0$-BSE calculations yield the A binding energies of 625 meV (MoS$_2$), 570 meV (MoSe$_2$), 580 meV (WS$_2$), and 530 meV (WSe$_2$), placing them toward the upper bounds of the experimental ranges. Some early theoretical works report binding energies exceeding 1 eV \cite{Eb_MoS2_MoSe2_WS2_WSe2_Rama-2012, Eb_MoS2_WS2_Shi-2013}, but these values often originate from simplified analytical models—such as Mott–Wannier model excitons—combined with GW-derived quasiparticle gaps, rather than full GW-BSE calculations. When the same studies employ a proper ab initio GW-BSE approach, the resulting binding energies align more closely with ours, emphasizing the critical role of explicitly treating the electron–hole interaction \cite{Eb_MoS2_MoSe2_WS2_WSe2_Rama-2012}.

\begin{table}[h!]
\centering
\caption{Unstrained exciton energies (in eV) for various monolayer TMDCs from this work and prior literature datasets. Here, TW, WE, and GW refer to the calculations performed in this work, those obtained from the Wannier model equation, and those computed using the GW-BSE method, respectively.}
\label{tab:strain_exciton-energies}
\begin{tabular}{l c c c c c c c}
\hline\hline
TMD&DX$_{\mathrm{KK}}$&IDX$_{\Gamma\mathrm{K}}$&IDX$_{\mathrm{K}\Lambda}$&IDX$_{\Gamma\Lambda}$&IDX$_{\mathrm{K'}\Lambda}$&MTD&REF  \\
\hline
\multirow{2}{*}{MoS$_2$} 
& 1.94 & 1.96 & 2.01 & 2.03 & 2.00 & TW & \\
& 1.84  & 1.91 & 1.90 & - & - & WE & \cite{emalik-1}  \\
& 2.16 &  -    & 2.18 & - & - & GW & \cite{Deilmann2019} \\

\hline
\multirow{2}{*}{MoSe$_2$} 
& 1.71 &  1.90 & 1.78 & 2.00 & 1.76 & TW & \\
& 1.51 &  1.73  & 1.65 & - & - & WE & \cite{emalik-1} \\
& 1.80 &  -    & 1.83 & - & - & GW & \cite{Deilmann2019} \\

\hline
\multirow{3}{*}{WS$_2$} 
& 2.08 & 2.22 & 1.96 & 2.11 & 1.94 & TW &\\
& 2.08 &  -    & 1.91 & - & - & EXP & \cite{Chand2022} \\
& 2.08 & 2.27 & 2.04 & 2.16 & - & WE & \cite{mos2-pl15} \\
& 1.99 & 2.14 & 1.98 & - & -  & WE & \cite{emalik-1}  \\
& 2.00 & 2.20 & 1.97 & - & –  & WE & \cite{Strain-1} \\
& 2.17 &  -    & 2.11 & - & - & GW & \cite{Deilmann2019} \\

\hline
\multirow{3}{*}{WSe$_2$} 
&  1.76 & 1.96 & 1.76 & 2.01 & 1.74 & TW & \\
& 1.67  & –    & 1.70 & - & –    & EXP & \cite{Strain-2}\\
& 1.74  & 2.17 & 1.69 & 2.02 & - &  WE & \cite{mos2-pl15}\\
& 1.65  & 2.05 & 1.59 & - & -    & WE & \cite{emalik-1} \\
& 1.78 &  -    & 1.72 & - & - & GW & \cite{Deilmann2019} \\
\hline\hline
\end{tabular}
\end{table}

In addition to excitonic states with zero center-of-mass momentum, we investigated finite-momentum excitons with $\mathbf{Q}$ = $\Lambda$, K, and M. For each momentum transfer, we analyzed the $k$-resolved exciton weights (as defined in the Section \ref{method}) and identified the lowest-energy excitonic states with dominant contributions corresponding to the $\mathrm{K}\Longleftrightarrow\Lambda$ (with momentum $\Lambda$), $\Gamma\Longleftrightarrow\Lambda$ (with momentum $\Lambda$), $\Gamma\Longleftrightarrow \mathrm{K}$ (with momentum K), and $\mathrm{K}^{\prime}\Longleftrightarrow\Lambda$ (with momentum M) transitions, as schematically illustrated in Fig.~\ref{fig-band}. The calculated energies for these indirect excitons (IDXs), labeled as IDX$_{\mathrm{K}\Lambda}$, IDX$_{\Gamma\Lambda}$, IDX$_{\Gamma \mathrm{K}}$, and IDX$_{\mathrm{K}^{\prime}\Lambda}$, are presented in Table \ref{tab:strain_exciton-energies}. 
The full $k$-space dispersion of these states—highlighting their characteristic multivalley structure—are provided in Fig.~S5 of the Supplemental Material.

To the best of our knowledge, only one experimental study has reported finite-momentum excitons for both WS$_2$ and WSe$_2$ \cite{Chand2022, Strain-2} monolayers. For WS$_2$, our results show good agreement, with an exact match for the direct exciton and a deviation of only $\sim$50 meV for IDX$_{\mathrm{K}\Lambda}$. For WSe$_2$, our calculations overestimate the DX$_{\mathrm{K}\mathrm{K}}$ and IDX$_{\mathrm{K}\Lambda}$ energies by approximately 90 and 60 meV, respectively. While the agreement is not fully quantitative, such differences may arise from the non-equivalent structural and dielectric environments considered in theory and experiment, including substrate screening, residual strain, and sample-specific structural conditions. In addition, the DFT-relaxed free-standing structures used in the calculations may not exactly reproduce the crystal configurations realized experimentally, where even small structural variations can noticeably shift excitonic energies.

Several theoretical values for excitonic energies have been reported, obtained either by solving the Wannier equation with a generalized Rytova–Keldysh screened Coulomb potential using strain-adapted band parameters from DFT calculations \cite{mos2-pl15, emalik-1, Strain-1}, or within the GW+BSE framework \cite{Deilmann2019}. When compared with these studies, as well as with the results obtained in the present work, clear deviations exist in the excitonic fingerprints—namely, the absolute energies of direct and indirect excitons and their relative energy separation, as seen in Table~\ref{tab:strain_exciton-energies}.  For the DFT band-structure-based analyses, these discrepancies most likely arise from the fact noted above: many-body corrections not only shift the absolute band gap but also modify the relative energy differences between the conduction- and valence-band valleys, which in turn directly influence the excitonic fingerprints.

\begin{figure}[t]
\includegraphics[width=\linewidth]{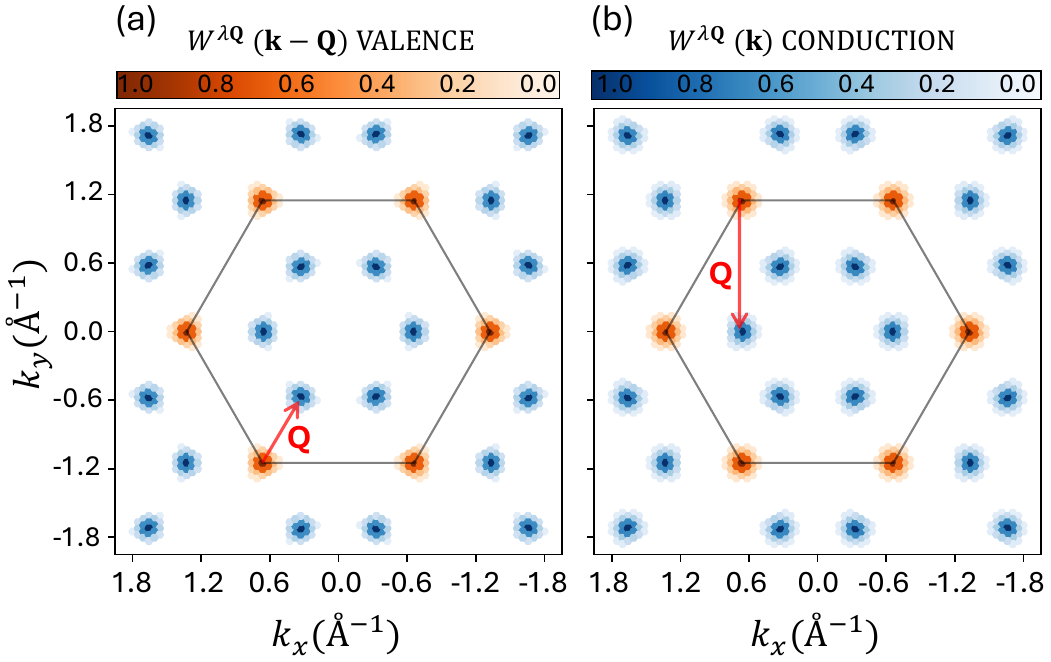}
\caption{\label{fig-x1} Calculated excitonic weights of MoS$_2$ for (a) IDX$_{\mathrm{K}\Lambda}$ and (b) IDX$_{\mathrm{K}^{\prime}\Lambda}$ (see Table~\ref{table3}), summed over all bands across the Brillouin zone (Eq.~\eqref{eq3}) and shown for every equivalent $\mathbf{Q}$-vector. The colour scales highlight the regions corresponding to the dominant electron-hole transitions for each chosen excitonic state and momentum.}
\end{figure}

For the indirect excitons, we further note the small energy difference between the IDX$_{\mathrm{K}\Lambda}$ and IDX$_{\mathrm{K}^{\prime}\Lambda}$ states (10-20 meV), which both involve transitions between the same bands, yet with different momentum. To clarify this behaviour, for the IDX$_{\mathrm{K}\Lambda}$ and IDX$_{\mathrm{K}^{\prime}\Lambda}$ excitons of MoS$_2$, we analyzed the excitonic weights, which quantify the dominant valence $|A_{cv}^{\lambda}(\mathbf{k}-\mathbf{Q})|^{2}$ and conduction $|A_{cv}^{\lambda}(\mathbf{k})|^{2}$ band contributions to each excitonic state (see Eq. 4). As shown in Fig. \ref{fig-x1}(a) and (b), both excitons, although located at different transfer momenta, are composed of the same electron and hole states. This strong similarity in their underlying band-structure contributions naturally explains the small energy splitting observed between the IDX$_{\mathrm{K}\Lambda}$ and IDX$_{\mathrm{K}^{\prime}\Lambda}$ excitons , which is then pred dominently determined by the $\mathbf{Q}$-dependence of the electron-hole interaction kernel $\Xi(\mathbf{Q})$ (see Eq.~\eqref{eq:ehkernel}). However, despite their similar electronic character, the temperature-dependent population and excitation dynamics of these intervalley excitons will differ significantly, both because their energy separation corresponds to an effective temperature difference of 100-200 K, and because phonon-mediated scattering rates for these indirect states will depend on lattice vibrations with different energies and momenta.

\begin{figure}[ht!]
\includegraphics[width=\linewidth]{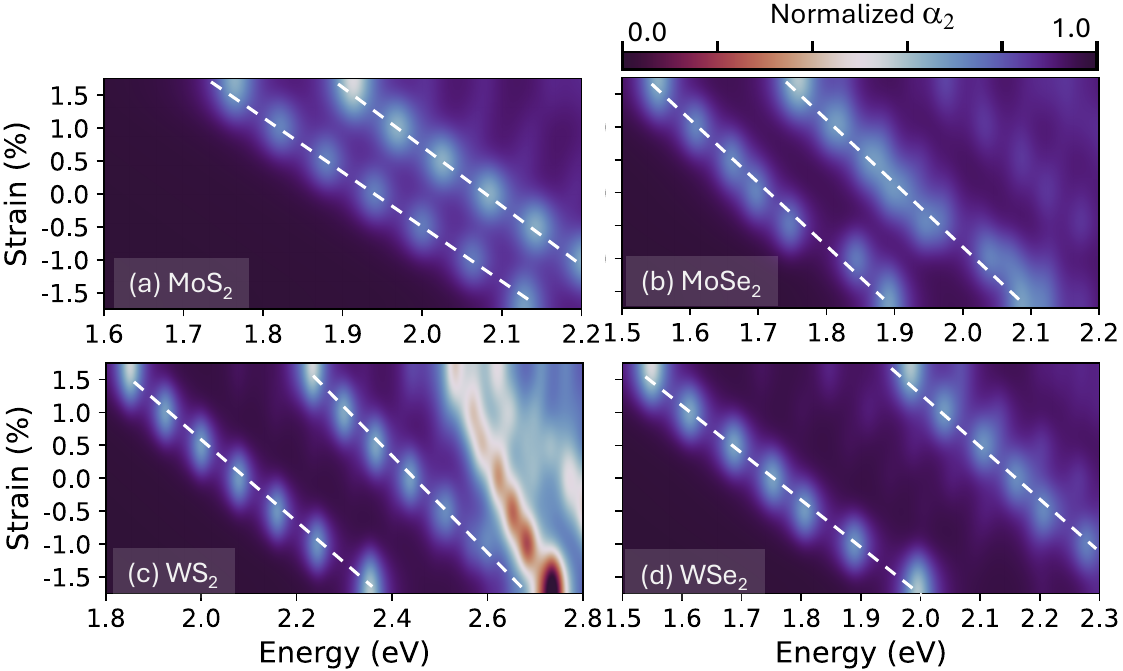}
\caption{\label{fig-3} Strain dependence of the direct exciton energies for (a) MoS$_2$, (b) MoSe$_2$, (c) WS$_2$, and (d) WSe$_2$. The colourmap indicates the evolution of the normalized 2D polarizability $\alpha_2$. The ground-state A and B excitonic transitions are highlighted, with white dashed lines denoting bright states and gray dashed lines marking dark spin-triplet states.}
\end{figure}

\subsection{Excitonic properties: Strained}
Building on the growing interest in strain engineering as an effective strategy to modulate excitonic properties in transition metal dichalcogenides, we present a systematic investigation of the strain dependence of exciton energies in MoS$_2$, MoSe$_2$, WS$_2$, and WSe$_2$. While previous studies have offered valuable insights into the strain response of individual excitonic states, our results provide a comprehensive comparative picture of strain-exciton coupling in the monolayer TMD materials family.

The colour maps in Fig.~\ref{fig-3} (a), (b), (c), and (d) display the evolution of the optical absorption spectra represented by the imaginary part of 2D polarizability, $\alpha_2$ from Eq.~\eqref{eq:alpha2D}, for monolayer MoS$_2$, MoSe$_2$, WS$_2$, and WSe$_2$, respectively. 
The dashed white lines trace the strain-induced shifts of the peak positions of the spin–orbit–split bright direct excitons, labeled A and B, corresponding to the DX$_{\mathrm{KK}}$ states. 
In addition, there are also the spin-triplet dark states (also observed experimentally\cite{Molas_2017}), which lie energetically below the bright singlet states.
With respect to the A-exciton energies, they are predicted to lie 16, 1, 40, and 24 meV below for MoS$_2$, MoSe$_2$, WS$_2$, and WSe$_2$, respectively. These dark states are in quantitative agreement with literature\cite{robert2021spin,Wang2018,Eb_WSe2_Shih-2024} and follow the same strain-induced trends as the bright excitons.

To quantitatively assess strain sensitivity, we evaluate the excitonic strain gauge factor-commonly employed in experimental characterization-, $\Omega = dE_X/d\varepsilon$, where $E_X$ is the corresponding strain dependent exciton energy and $\varepsilon$ is the strain (expressed in meV / strain percentage). The resulting values for A and B exciton are summarized in Table \ref{tab:strain_gauge}. The values for A-exciton are systematically larger than those obtained from the Wannier equation using DFT-derived effective masses and band edges \cite{emalik-1}. However, for W-based compounds, calculations based on a similar Wannier–Keldysh approach \cite{mos2-pl15} yield results that are comparable with ours. The pronounced redshift in the energy of the A-exciton under applied tensile strain, as predicted for all the considered monolayers, is consistent with photoluminescence (PL) measurements \cite{mos2-pl15, emalik-2}. The experimental strain gauge factors for A-exciton shifts are reported as –102 meV/\% for WS$_2$ and –118 meV/\% and –46 meV/\% for WSe$_2$. Our calculations yield -164 meV/\% for WS$_2$ and -143 meV/\% for WSe$_2$, systematically exceeding the experimental values. These quantitative differences likely reflect the absence of a direct one-to-one correspondence between the ideal biaxial strain considered theoretically and the actual strain transferred to experimental samples, together with additional effects associated with substrate screening, dielectric environment, temperature, and sample-specific structural conditions.

\begin{table}[t]
\caption{\label{tab:strain_gauge} Strain gauge factors (in meV/\%) for direct (DX$_{\mathrm{K}\mathrm{K}}$) and indirect (IDX$_{\Gamma K}$, IDX$_{\mathrm{K}\Lambda}$) excitons in strained monolayer TMDs. Here, TW, WE, and EXP refer to the calculations performed in this work, those obtained from the Wannier equation, and those measured experimentally, respectively. }
\centering
\scalebox{0.88}{
\begin{tabular}{l c c c c c c c c c}
\hline\hline
TMD  & A & B & IDX$_{\Gamma \mathrm{K}}$ & IDX$_{\mathrm{K}\Lambda}$ & IDX$_{\Gamma\Lambda}$ & IDX$_{\mathrm{K}'\Lambda}$ & IDX$_{\mathrm{K}\mathrm{K}'}$ & MTD & Ref.  \\\hline
\multirow{2}{*}{MoS$_2$}
 & -121 & -118 & -233 & 37 & -81 & 36 & -120 & TW\\
 & -72 & - & -142 & 10.1 & - & - & - & WE & \cite{emalik-1}\\\hline
\multirow{2}{*}{MoSe$_2$}
 & -112 & -111 & -201 & 34 & -75 & 35 & -111 & TW \\
 & -65 & - & -123 & 8 & - & - & - & WE & \cite{emalik-1}\\\hline
\multirow{5}{*}{WS$_2$}
 & -164 & -160 & -261 & 27 & -90 & 28 & -156 & TW \\
 & -102 & - & - & - & -68 & - & - & EXP & \cite{mos2-pl15}\\
 & -60 & - & -123 & 37 & - & - & - & WE & \cite{emalik-1}\\
 & -113 & - & -236 & 74 & - & - & - & WE & \cite{Strain-1}\\
 & -125 & & -223 & 45 & -52 & - & - & WE & \cite{mos2-pl15}  \\
\hline
\multirow{5}{*}{WSe$_2$}
 & -143 & -122 & -228 & 18 & -73 & 18 & -139 & TW \\
 & -46 & - & - & 22 & - & - & - & EXP &\cite{Strain-2}\\
 & -118 & - & - & 34 & - & - & - & EXP & \cite{mos2-pl15}\\
 & -54 & - & -106 & 37 & - & - & - & WE & \cite{emalik-1}\\
 & -111 & - & -197 & 43 & -51 & - & - & WE &\cite{mos2-pl15}\\\hline\hline
\end{tabular}
}
\end{table}

To elucidate more the origin of these shifts, we compare them with the strain-induced variations of the direct band gap at the K point, obtained from G$_0$W$_0$-corrected band-gap values (Fig.~\ref{fig-strain}). The calculated gauge factors for the direct exciton energy and the corresponding band gap gauge factors depicted in Table \ref{tab:strain_gap} are close to eacother, indicating that the change in band gap is the primary driver of the direct exciton redshift under strain.

\begin{table}[b]
\caption{\label{tab:strain_gap} Calculated strain gauge factors (in meV/\%) for the band gaps ($\Delta E$) associated with direct and indirect exciton formation.}
\centering
\begin{tabular}{l c c c c c}
\hline\hline
Material & $\Delta E_{\mathrm{KK}}$ & $\Delta E_{\mathrm{K}\Gamma}$ & $\Delta E_{\mathrm{K}\Lambda}$ & $\Delta E_{\Gamma\Lambda}$ & Calc. \\\hline
\multirow{2}{*}{MoS$_2$} & -137 & -266 & 30 & -99 & G$_0$W$_0$\\
& -109 & -233 & 39 & -84 & PBE\\\hline
\multirow{2}{*}{MoSe$_2$} & -126 & -235 & 29 & -80 & G$_0$W$_0$\\
& -84 & -157 & 5 & -69 & PBE\\\hline
\multirow{2}{*}{WS$_2$} & -170 & -283 & 12 & -101 & G$_0$W$_0$ \\
 &-133 & -253 & 35 & -85 & PBE \\\hline
\multirow{2}{*}{WSe$_2$} & -149 & -248 & 13 & -86 & G$_0$W$_0$\\
 & -96 & -159 & -8 & -71 & PBE\\
\hline\hline
\end{tabular}
\end{table}

To gain insight into the strain response of the indirect excitonic manifold, we analyze the structure of the excitonic wave functions, by examining the lowest-energy indirect excitons corresponding to the most important valleys, as discussed above, also including the higher energy states for each valley: in particular, we analyze the first ten states for IDX$(\mathbf{Q}=\Lambda)$, IDX$(\mathbf{Q}=\mathrm{K})$, and IDX$(\mathbf{Q}=\mathrm{M})$, as shown in Figs. S6-S10 in SM.

\begin{figure}[t]
\includegraphics[width=\linewidth]{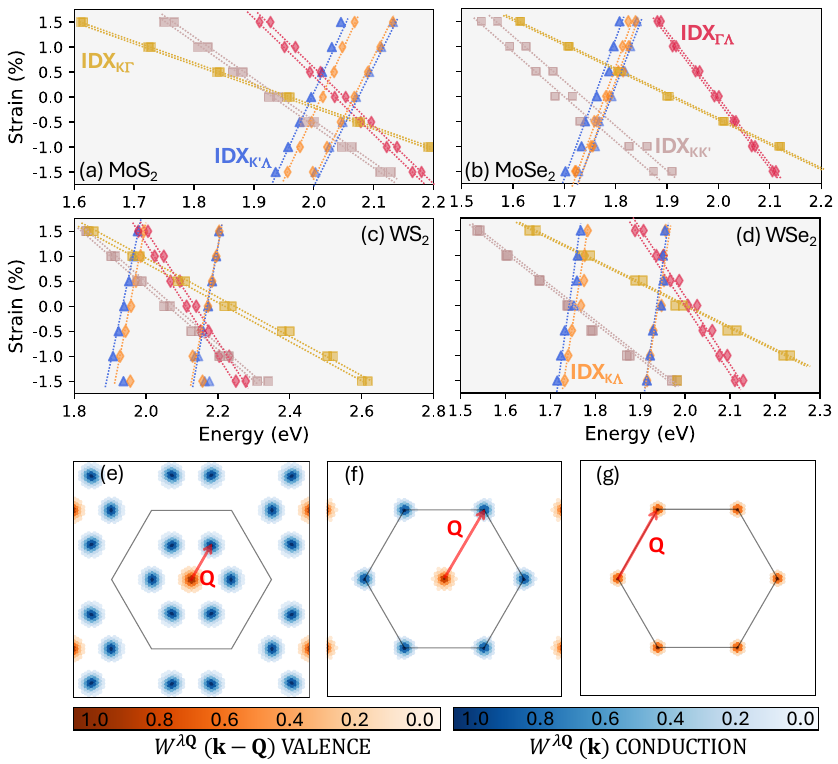}
\caption{\label{fig-XII} Strain dependence of the energies of the indirect excitons,IDX$_{\mathrm{K^{\prime}\Lambda}}$ (blue triangles), IDX$_{\mathrm{K\Lambda}}$ (orange diamonds), IDX$_{\Gamma\Lambda}$ (red diamonds), IDX$_{\Gamma \mathrm{K}}$ (yellow squares), and IDX$_{\mathrm{KK^{\prime}}}$ (rose squares), for (a) MoS$_2$, (b) MoSe$2$, (c) WS$2$, and (d) WSe$2$. The corresponding excitonic weight distributions for (e) DX$_{\Gamma\Lambda}$, (f) IDX$_{\mathrm{\Gamma K}}$, and (g) IDX$_{\mathrm{KK^{\prime}}}$ are also shown, similarly to Fig.~\ref{fig-x1}. The colour maps represent the normalized weights on a linear scale. } 
\end{figure}

The linear blue shift of the IDX$(\mathbf{Q}=\mathrm{M})$ excitons is consistently observed across all considered materials, indicating that excitons within the examined energy window possess nearly identical gauge factors. To further substantiate this behaviour, we focus on the strain evolution of the ground-state (degenerate) exciton and the next excited state, highlighted by blue triangles in Figs.~\ref{fig-XII}(a-d). These excitons exhibit the characteristic IDX$_{\mathrm{K^{\prime}}\Lambda}$ weight distribution shown in Fig.~\ref{fig-x1}(b), with only minor strain-induced modifications. As expected for excitons dominated by a single valley pair, their strain dependence closely follows the evolution of the underlying indirect conduction–valence band gap, as presented in Fig.~S10 of the SM.

In the unstrained case, these IDX$_{\mathrm{K^{\prime}}\Lambda}$ excitons lie at higher energies than the direct DX$_{\mathrm{K}\mathrm{K}}$ exciton in Mo-based systems (see rose squares for comparison), whereas they appear at lower energies in W-based systems. However, owing to the opposite strain response of these indirect excitonic states compared to direct excitons, several IDX$_{\mathrm{K^{\prime}}\Lambda}$ excitons become energetically favourable and fall well below the A-exciton for all materials already at a compressive strain of 1.5\%.

The ID$(\mathbf{Q}=\Lambda)$ excitons, indicated by red diamonds in Figs.~\ref{fig-XII}(a-d), exhibit a more intricate strain response. Unlike excitons at other center-of-mass momenta, two distinct groups with opposite strain dependencies are clearly resolved, particularly in MoS$_2$ and WS$_2$. A systematic analysis of the excitonic weight distributions elucidates the microscopic origin of this behaviour. Specifically, two different types of electronic transitions to the conduction $\Lambda$ valley, either from $\Gamma$ or from K in valence, contribute to the low-energy $\mathbf{Q}=\boldsymbol{\Lambda}$ excitonic valley. At higher energies, this gives rise to hybridized states with a broad momentum distribution that is not strongly localized around the $\Gamma$, $\Lambda$ or K points. To avoid ambiguities arising from such hybridization, we focus on the lowest-energy exciton each type, which is strongly localized: the IDX$_{\mathrm{K}\Lambda}$ exciton weight distribution was already presented in Fig.~\ref{fig-x1}(a), while the IDX$_{\Gamma\Lambda}$ states are shown in Fig.~\ref{fig-XII}(e).

The predicted opposite intravalley gauge factors for IDX$_{\Gamma\Lambda}$ and IDX$_{\mathrm{K}\Lambda}$ are consistent with the corresponding strain-dependent band-gap trends shown in Fig.~S10. Moreover, the nearly identical gauge factors obtained for IDX$_{\mathrm{K}\Lambda}$ and IDX$_{\mathrm{K^{\prime}}\Lambda}$ are naturally explained by the fact that their underlying band-gap trend is the same. Our calculations further demonstrate that IDX$_{\Gamma\Lambda}$ excitons always remain at higher energies than the direct DX$_{\mathrm{KK}}$ exciton, whereas IDX$_{\mathrm{K}\Lambda}$, analogous to IDX$_{\mathrm{K^{\prime}}\Lambda}$, becomes a well-separated low-energy state under compressive strain.

For IDX$(\mathbf{Q}=\mathrm{K})$ excitons, a clear linear strain dependence is observed, most prominently in WSe$_2$ (and similarly in WS$_2$ and MoSe$_2$). However, analogous to the $\mathbf{Q}=\Lambda$ case, band and valley hybridization complicate a unique assignment of excitonic character across the full strain range. We therefore focus on the two most localized low-energy excitonic transitions, namely the K–K$^{\prime}$ and $\Gamma$-K configurations, denoted as IDX$_{\mathrm{\Gamma K}}$ and IDX$_{\mathrm{KK^{\prime}}}$ respectively, as illustrated in Figs.~5(f) and 5(g).

The strain dependence of IDX$_{\mathrm{KK^{\prime}}}$ (rose squares) closely mirrors that of the direct DX$_{\mathrm{KK}}$ exciton, as expected since they originate from the same electron-hole wavefunctions and from their nearly identical $k$-space weight distributions. In contrast, the K–$\Gamma$ localized exciton IDX$_{\mathrm{\Gamma K}}$ (yellow squares) exhibits a noticeably larger red-shift gauge factor and becomes energetically lower than the A-exciton in MoS$_2$ and WS$_2$ already at 1.5\% tensile strain.

The strain gauge factors extracted for the identified excitons are summarized in Table~\ref{tab:strain_gauge}. Overall, the qualitative trends are consistent with those obtained from Wannier-Keldysh approaches, although noticeable quantitative deviations remain. In particular, the values reported by Khatibi \textit{et al.} differ substantially from our predictions, whereas the gauge factors computed by Kumar \textit{et al.} for IDX$_{\Gamma \mathrm{K}}$ are in close agreement with our results. For the IDX$_{\Gamma\Lambda}$ exciton, Kumar \textit{et al.}~\cite{mos2-pl15} experimentally reported a gauge factor of –68 meV/\% in monolayer WS$_2$, and a positive gauge factor of +34 meV/\% for IDX$_{\mathrm{K}\Lambda}$ in WSe$_2$, opposite in sign to IDX$_{\Gamma\Lambda}$ in WS$_2$. In our calculations, these values are obtained as –90 and +18 meV/\%, respectively, reproducing the experimentally observed qualitative behaviour. For IDX$_{\mathrm{K}\Lambda}$ in WSe$_2$, Aslan \textit{et al.}~\cite{Strain-2} likewise confirmed a positive gauge factor of +22 meV/\%, again consistent with our findings. We note, however, that the same study reported a DX$_{\mathrm{KK}}$ gauge factor that differs significantly from both our results and those of Kumar \textit{et al.} The opposite strain responses of IDX$_{\Gamma\Lambda}$ and IDX$_{\mathrm{K}\Lambda}$ were further corroborated by Wannier–Keldysh calculations in Ref.~\cite{mos2-pl15}, yielding –52 and +45 meV/\% for WS$_2$ and –51 and +43 meV/\% for WSe$_2$, respectively. 
In summary, while full quantitative agreement with the experimental literature has not yet been achieved, our results capture the correct qualitative behaviour and provide a more detailed microscopic understanding of the origin of these strain-induced exciton shifts.

Going back to the level of single-particle band structure, we computed the indirect band-gap shifts, as previously done for the direct case, by identifying the valence and conduction band edges that that correspond to the main exciton valleys (Table~\ref{tab:strain_gap}). As already remarked, the resulting G$_0$W$_0$ gauge factors show clear correspondence with those extracted for the associated excitons. Furthermore, the opposite strain responses of IDX$_{\Gamma\Lambda}$ and IDX$_{\mathrm{K}\Lambda}$, also observed experimentally, closely follows the behaviour of their respective band edges, reinforcing the direct link between valley-specific band-structure modifications and excitonic shifts. The discrepancies between the gauge factors obtained at the G$_0$W$_0$ and PBE levels are substantial. Given the near one-to-one correspondence between band-edge shifts and exciton energies, these differences underscore the necessity of many-body–corrected band structures when modeling strain responses within simplified excitonic frameworks. Indeed, even small energy variations in the excitonic picture can significantly affect the interpretation of experimental observations.

\section{Conclusion}
We have investigated the strain-dependent excitonic landscape of monolayer MoS$_2$, MoSe$_2$, WS$_2$, and WSe$_2$ using fully relativistic G$_0$W$_0$+BSE calculations, including both zero- and finite-center-of-mass-momentum excitons. By resolving the reciprocal-space composition of the BSE eigenstates, we identify the valley character of the low-energy excitonic states and follow their evolution under biaxial strain.

A central result of this work is that strain does not produce a uniform shift of the excitonic spectrum, but instead reorganizes the low-energy manifold in a strongly valley-dependent manner. Excitons originating from different combinations of the K, $\Gamma$, and $\Lambda$ valleys exhibit markedly different, and in several cases opposite, strain responses. As the relative valley energies evolve, the energetic ordering of direct and finite-momentum excitons can change, while states involving competing valley configurations may acquire substantial hybridization. These effects demonstrate that the strain response of monolayer TMDs cannot be characterized solely through the evolution of the bright K–K exciton.

The calculated excitonic strain gauge factors closely follow the strain dependence of the corresponding G$_0$W$_0$ quasiparticle gaps, establishing a direct microscopic connection between valley-specific band-structure reconstruction and the resulting excitonic shifts. At the same time, the substantial differences between PBE and G$_0$W$_0$ valley offsets and strain derivatives show that many-body corrections are important not only for absolute excitation energies but also for predicting the relative ordering and strain response of finite-momentum excitons. The calculated trends are consistent with the experimentally observed valley-dependent strain behaviour where comparison is available. Overall, these results provide a many-body framework for understanding and controlling the multivalley excitonic landscape of monolayer TMDs through strain.

\section{Methods\label{method}}
The single-particle wavefunctions and corresponding energies used as input for the G$_0$W$_0$ and Bethe–Salpeter Equation (BSE) calculations were obtained from Kohn–Sham Density Functional Theory (DFT), as implemented in Quantum ESPRESSO (QE) \cite{giannozzi2020quantum}. These calculations employed fully relativistic, norm-conserving pseudopotentials within the generalized gradient approximation (GGA), using the Perdew–Burke–Ernzerhof (PBE) exchange–correlation functional \cite{perdew1996generalized, hamann2013optimized}. The pseudopotentials were sourced from the PseudoDojo project \cite{van2018pseudodojo}.

The pseudopotentials included semi-core electrons, with valence configurations of 4$s^2$4$p^6$5$s^1$4$d^5$ for molybdenum, 5$s^2$5$p^6$6$s^2$5$d^4$ for tungsten, 3$s^2$3$p^4$ for sulfur, and 3$d^{10}$4$s^2$4$p^4$ for selenium. Spin-orbit coupling (SOC) effects were fully included through spinorial wavefunctions. A plane-wave energy cutoff of 120 Ry was used, along with a symmetric Monkhorst–Pack $k$-point grid of (42$\times$42$\times$1) (169 $k$-points in the irreducible Brillouin zone, centered at $\Gamma$) for all materials.

To eliminate spurious interlayer interactions, a vacuum layer of 60 Bohr was added along the out-of-plane direction, and a 2D Coulomb truncation scheme was applied \cite{sohier2017density}. van der Waals (vdW) interactions were incorporated using Grimme’s D2 dispersion correction \cite{grimme2006semiempirical, barone2009role}. All calculations were repeated for biaxial strain values ranging from -1.5\% to +1.5\%, keeping the computational parameters fixed. For each strain value, full ionic relaxation was included.

The Many-Body Perturbation Theory (MBPT) calculations were performed using the YAMBO code \cite{marini2009yambo, sangalli2019many}, with the DFT wavefunctions and energies as input. A 2D slab geometry was employed for truncating Coulomb interactions \cite{Rozzi2006, guandalini2023efficient}. The G$_0$W$_0$ corrections to the quasiparticle energies were calculated using the Godby–Needs plasmon-pole approximation for the dynamically screened interaction \cite{Godby1989, Oschlies1995}. Convergence of both direct and indirect band gaps was ensured ($<$ 10 meV in energy of A and B excitons) using 500 bands for the screened Coulomb potential $W$ and 600 bands for the Green’s function $G$, with screening computed within the Random Phase Approximation (RPA). These corrections were applied to the top four valence bands and the bottom four conduction bands. The Bethe–Salpeter Equation (BSE) for excitonic states was solved within the Tamm–Dancoff approximation (uncoupled resonant and antiresonant transitions), \cite{Strinati1988, smith2017interacting}. The details regarding the converged computational parameters used in Yambo calculations can be found in the Supplemental Materials (SM).

The BSE two-particles Hamiltonian, $H^{BSE}(\mathbf{Q})$, describes the coupling of single-particle transitions with momentum difference $\mathbf{k}_1 - \mathbf{k}_2 = \mathbf{Q}$, mediated by electron–hole Coulomb interactions. These interactions are approximated as the sum of the direct Hartree (H) and statically screened exchange (SEX) terms \cite{Martin2016}:
\begin{eqnarray}
H^{\textrm{BSE}}_{\substack{ c_1 v_1 \mathbf{k}_1 \\ c_2 v_2 \mathbf{k}_2}}(\mathbf{Q}) &=& (\epsilon_{c_1\mathbf{k}_1} - \epsilon_{v_1\mathbf{k}_1-\mathbf{Q}})\delta_{\substack{ c_1 v_1 \mathbf{k}_1 \\ c_2 v_2 \mathbf{k}_2}}\nonumber\\ &+& (f_{v_2 \mathbf{k}_2}-f_{c_2 \mathbf{k}_2})\Xi^{\textrm{HSEX}}_{\substack{ c_1 v_1 \mathbf{k}_1 \\ c_2 v_2 \mathbf{k}_2}}(\mathbf{Q}).
\end{eqnarray}
Here, $\epsilon_{nk}$ refers to a quasiparticle energy in the GW approximation, with occupation number $f_{nk}$.

Accordingly, the electron-hole kernel is composed by two contributions coming from the two pieces of the interaction:
\begin{equation}\label{eq:ehkernel}
    \Xi^{\textrm{HSEX}}_{\substack{ c_1 v_1 \mathbf{k}_1 \\ c_2 v_2 \mathbf{k}_2}}(\mathbf{Q}) =  -W^{\textrm{SEX}}_{\substack{ c_1 v_1 \mathbf{k}_1 \\ c_2 v_2 \mathbf{k}_2}}(\mathbf{Q}) + V^{\textrm{H}}_{\substack{ c_1 v_1 \mathbf{k}_1 \\ c_2 v_2 \mathbf{k}_2}}(\mathbf{Q}),
\end{equation}
with $W^{\textrm{SEX}}$ representing the attractive screened interaction (responsible for the exciton binding) and $V^{\textrm{H}}$ being the repulsive, unscreened exchange interaction. The latter term is zero for triplet excitons because of how it couples electronic wavefunctions (singlet excitons are the only ones that may be excited optically).

Exciton energies $E_{\lambda}({\mathbf{Q}})$ and corresponding wavefunctions were obtained by diagonalizing the BSE Hamiltonian for each center-of-mass (COM) momentum $\mathbf{Q}$:
\begin{equation}
\sum_{c_2 v_2 \mathbf{k}_2} H^{\textrm{BSE}}_{\substack{ c_1 v_1 \mathbf{k}_1 \\ c_2 v_2 \mathbf{k}_2}}(\mathbf{Q}) A^{\lambda \mathbf{Q}}_{c_2 v_2 \mathbf{k}_2} = E_{\lambda}({\mathbf{Q}}) A^{\lambda \mathbf{Q}}_{c_1 v_1 \mathbf{k}_1}.
\end{equation}
Here, $A^{\lambda \mathbf{Q}}_{cv\mathbf{k}} = \langle c\mathbf{k} \otimes v\mathbf{k}-\mathbf{Q}| \lambda \mathbf{Q}\rangle$ represent the projections of the excitonic eigenvectors onto the electron–hole product states for a transition from valence band $v$ with momentum $\mathbf{k}-\mathbf{Q}$ to a conduction band $c$ with momentum $\mathbf{k}$. These coefficients describe the transformation from the single-particle transition basis to the excitonic basis in which $H^{\textrm{BSE}}(\mathbf{Q})$ is diagonal. As demonstrated in Sec. \ref{sec:BSEresults}, the dominant influence of strain stems from modifications to the single-particle states $|c \mathbf{k}\rangle$ and $|v \mathbf{k}\rangle$.

To characterize the momentum-space distribution of excitonic states, we analyzed the $k$-resolved exciton weights at a chosen state $\lambda$ and momentum $\mathbf{Q}$, defined as, 
\begin{equation}
    \mathcal{W}^{\lambda \mathbf{Q}}(\mathbf{k}) = \sum_{vc}\left|A^{\lambda \mathbf{Q}}_{vc\mathbf{k}}\right|^2, \label{eq3}
\end{equation}
providing a direct measure of the exciton's distribution across the Brillouin zone. Peaks in $\mathcal{W}^\lambda(\mathbf{k})$ indicate the dominant regions in reciprocal space contributing to a given excitonic state, enabling classification according to their character and transfer momentum, such as $K$–$K$ ($\mathbf{Q}=0$), $K$–$\Lambda$ ($\mathbf{Q}=\boldsymbol{\Lambda}$), or $\Gamma$–$K$ ($\mathbf{Q}=\mathbf{K}$).

Moreover, we define the exciton binding energy as the difference between the exciton energy and the corresponding minimum quasiparticle gap \textit{at the same momentum transfer vector:}
\begin{equation}
    E_b(\lambda;\mathbf{Q}) =  \min_{vc\mathbf{k}}\{\epsilon_{c\mathbf{k}}-\epsilon_{v\mathbf{k}-\mathbf{Q}}\} - E_\lambda (\mathbf{Q})
\end{equation}

Finally, the optical absorption spectrum including excitonic effects is given by the imaginary part of the 2D polarizability $\alpha(\omega)$ (which is related to the macroscopic dielectric function\cite{Cudazzo2011}) computed at the BSE level:
\begin{equation}\label{eq:alpha2D}
    \alpha_{2}(\omega)=\frac{2\pi}{A}\sum_{\lambda} \left|\boldsymbol{\varepsilon}\cdot\sum_{cv\mathbf{k}}A^{\lambda \mathbf{0}}_{cv\mathbf{k}}  \mathbf{d}_{cv\mathbf{k}}\right|
^2 \delta_\eta (\omega - E_\lambda (\mathbf{0})).
\end{equation}
Here, $\mathbf{d}_{cv\mathbf{k}}$ is the single-particle interband dipole matrix element, $\boldsymbol{\varepsilon}$ the polarization direction of the incoming light, and $A$ the 2D unit cell area. $\eta$ is a numerical broadening parameter expanding the delta functions into Lorentzian peaks. 

\section*{Supplementary Material}
The Supplemental Material includes the following additional results: Yambo input parameters, calculated band structures for all studied materials under all considered strain values; the exciton weight distribution in the Brillouin zone for indirect (finite-momentum) excitons; strain-induced variations in direct and indirect exciton energies; and strain-dependent changes in direct and indirect GW quasiparticle band gaps.

\section*{Acknowledgements}
This work was supported by the Research Foundation-Flanders (FWO) and the Special Research Funds of the University of Antwerp (BOF-UA). We acknowledge EuroHPC Joint Undertaking for awarding the EHPC-EXT-2024E02-090 access to Leonardo-Booster@Cineca, Italy. F.P. acknowledges funding by ICSC - Centro Nazionale di Ricerca in High Performance Computing, Big Data and Quantum Computing – funded by the European Union through the Italian Ministry of University and Research under PNRR M4C2I1.4 (Grant No. CN00000013).

\section*{data availability statement}
The data that support the findings of this study are openly available in Zenodo at http://doi.org/, reference number [reference number will be provided after acceptance].

\bibliography{apssamp}

\end{document}